\documentclass[a4paper,fleqn,usenatbib]{mnras}

\usepackage{newtxtext,newtxmath}

\usepackage[T1]{fontenc}
\usepackage{ae,aecompl}

\usepackage{graphicx}	
\usepackage{amsmath}	

\newcommand{\nustar}{\textit{NuSTAR}~}
\newcommand{\swift}{\textit{Swift}~}
\newcommand{\eps}{erg s$^{-1}$~}

\newcommand{\ecps}{erg cm s$^{-1}$~}
\newcommand{\pcm}{cm$^{-2}$~}
\newcommand{\M}{$M_{\odot}$~}
\newcommand{\phc}{ph cm$^{-2}$ s$^{-1}$}

\newcommand{\nh}{$N_{\rm H}$~}
\newcommand{\gps}{gm s$^{-1}$~}

\title[Extragalactic BHXRBs]{{\it NuSTAR} and {\it Swift}  Observations of the Extragalactic  Black Hole X-ray Binaries}

\author[Jana et al.]{
Arghajit Jana$^{1}$\thanks{E-mail: argha@prl.res.in, argha0004@gmail.com},
Sachindra Naik$^{1}$,
Debjit Chatterjee$^{2}$,
Gaurava K. Jaisawal$^{3}$
\\
$^{1}$Astronomy \& Astrophysics Division, Physical Research Laboratory, Navrangpura, Ahmedabad, 380009, India\\
$^{2}$Indian Institute of Astrophysics, Koramangala, Bangalore 560034, India \\
$^{3}$National Space Institute, Technical University of Denmark, Elektrovej, 327-328, DK-2800 Lyngby, Denmark \\
 }
\date{Accepted XXX. Received YYY; in original form ZZZ}

\pubyear{2020}

\begin{document}

\label{firstpage}
\pagerange{\pageref{firstpage}--\pageref{lastpage}}
\maketitle

\begin{abstract}
We present the results obtained from detailed spectral and timing studies of extra-galactic black hole X-ray binaries LMC~X--1 and LMC~X--3, using simultaneous observations with {\it Nuclear Spectroscopic Telescope Array (NuSTAR)} and {\it Neil Gehrels Swift} observatories. The combined spectra in the $0.5-30$~keV energy range, obtained between 2014 and 2019, are investigated for both sources. We do not find any noticeable variability in $0.5-30$~keV light curves, with $0.1-10$~Hz fractional rms estimated to be $<2$\%. No evidence of quasi-periodic oscillations is found in the power density spectra. The sources are found to be in the high soft state during the observations with disc temperature $T_{\rm in}\sim 1$~keV, photon index, $\Gamma > 2.5$ and thermal emission fraction, $f_{\rm disc}>80$\%. An Fe K$\alpha$ emission line is detected in the spectra of LMC~X--1, though no such feature is observed in the spectra of LMC~X--3. From the spectral modelling, the spins of the black holes in LMC~X--1 and LMC~X--3 are estimated to be in the range of $0.92-0.95$ and $0.19-0.29$, respectively. The accretion efficiency is found to be, $\eta \sim 0.13$ and $\eta \sim 0.04$ for LMC~X--1 and LMC~X--3, respectively.
\end{abstract}

\begin{keywords}
X-Rays:binaries -- stars: individual: (LMC~X--1, LMC~X--3) -- stars:black holes -- accretion, accretion discs
\end{keywords}


\section{Introduction}
\label{sec:intro}
A stellar-mass black hole X-ray binary (BHXB) consists of a primary black hole of mass $M_{\rm BH} \sim 3-100$ \M and a normal companion star. A BHXB can be classified as a high mass black hole X-ray binary (HMBHXB) or low mass black hole X-ray binary (LMBHXB) depending on the mass of the companion star \citep{RM06,Tetarenko2016}. An HMBHXB consists of an O, B, or WR star, while an LMBHXB consists of an A-type or later star as a companion. Depending on the X-ray activity, a BHXB is classified as a persistent or transient source. A persistent source appears to be always active in X-rays with luminosity, $L_{\rm X} \geq 10^{35}$ \eps and easily detectable with the instruments on-board X-ray observatories. On the other hand, a transient source spends most of the time at the quiescent state ($L_{\rm X} \sim 10^{30-32}$ \eps) with occasional outbursts when the X-ray luminosity increases to $L_{\rm X} \sim 10^{35-38}$ \eps.

The BHXBs show rapid variation and fluctuations in the spectral and timing properties. The X-ray spectrum of the source in these binaries can be described by a model consisting of a multi-colour thermal blackbody and a non-thermal hard power-law tail. The thermal multi-colour blackbody component is believed to originate from the standard thin disk \citep{SS73,NT73}, whereas the hard power-law tail is believed to be originated by the up-scattering of blackbody photons in the Compton cloud located close to the black hole \citep{ST80,ST85,HM93,Z93,CT95,Done2007}. During the outburst period, a BHXB goes through different spectral states, namely, low hard state (LHS), hard-intermediate state (HIMS), soft-intermediate state (SIMS), and high soft state (HSS) \citep{RM06,AJ2020b}. The LHS is dominated by the Comptonized hard photons with the power-law photon index $\Gamma \sim 1.5-1.7$ and disk temperature, $T_{\rm in}\sim 0.4-0.5$~keV, while the HSS is dominated by the thermal soft photons with $\Gamma \gtrsim 2.5$ and $T_{\rm in} \sim 1$~keV \citep[for details, see][]{RM06,AJ2021b}.

To date, over 100 stellar mass BHXBs have been discovered \citep{Tetarenko2016}. Among the known population, only five BHXBs, namely, LMC~X--1, LMC~X--3, IC~10~X--1, M33~X-7, and NGC~300~X--1 have an extra-galactic origin. Of these five sources, LMC~X--1 and LMC~X--3 are nearby sources and are located at a distance of $\sim 48.1$~kpc in the Large Magellanic Cloud (LMC), a satellite galaxy of Milky Way \citep{Shattow2009}. LMC~X--1 and LMC X--3 are the second and third black holes discovered after Cygnus~X--1 \citep{Cowley1983,Hutchings1983}.

LMC~X--1 is the first extra-galactic X-ray source discovered by UHURU \citep{Mark1969,Price1971}. It is a persistent X-ray source in an HMBHXB system containing an optical giant OIII star of mass $M_{*}=31.79 \pm 3.48$ $M_{\odot}$ \citep{Orosz2009}. The black hole (BH) mass in this binary system is found to be $M_{\rm BH} = 10.91\pm1.41$ $M_{\odot}$ \citep{Orosz2009}. The orbital period, distance and the inclination angle of the binary system are reported to be $P_{\rm orb} \sim 3.909 \pm 0.0005$~days, $d \sim 48.1$~kpc and $i = 36^{\circ{}}.38 \pm 1.^{\circ{}}92$, respectively \citep{Orosz2009}. The BH is found to be rapidly rotating with spin parameter of $a^* \geq 0.92$ \citep{Gou2009,Mudambi2020,Bhuvana2021}. LMC~X--1 does not show any long-term variability, though short-term variabilities have been reported by \citet{Nowak2001}. Consistent with the long-term light curve, the source is always found to be in the thermal dominated spectral state \citep{Nowak2001}. The power density spectra are similar to that of a high soft state (HSS) and can be described with a power-law model \citep{Bhuvana2021}.

LMC~X--3 is a persistent HMBHXB consisting a BH of mass $M_{\rm BH} = 6.98 \pm 0.56$ $M_{\odot}$, and a B3~V companion star of mass, $M_{*} = 3.63 \pm 0.57$ $M_{\odot}$ \citep{Orosz2014}. The orbital period, distance and the inclination angle of the system are reported to be, $P_{\rm orb} \sim 1.7$~days, $d\sim 48.1$~kpc and $i=69^{\circ{}}.24 \pm 0.72^{\circ{}}$, respectively \citep{Orosz2014}. The spin of the BH in LMC~X--3 is reported to be low with $a^* \leq 0.3$ \citep{Steiner2010,Steiner2014,Bhuvana2021}. LMC~X--3 is found to spend most of the time in thermal dominated state with occasional transition to the LHS \citep{Boyd2000,Wilms2001,Smale2012}. Occasionally, LMC~X--3 is observed to be in anomalous low/hard state (ALS) with X-ray intensity reduced to $L_{\rm X} \sim 10^{35}$ \eps \citep{Torpin2017}.

In this paper, we present the results of spectral and temporal analysis of extra-galactic BHXBs LMC~X--1 and LMC~X--3, using data from the \nustar and \swift observations over a duration between 2014 and 2019. The paper is organized in the following way. In \S2, we present the details of observations and data reduction processes. In \S3, we present the result of timing and spectral analysis. In \S4, we discuss our results and in \S5, we summarize our findings from the study.

\begin{table*}
\caption{Log of observations of LMC~X--1 and LMC~X--3.}
\label{tab:log}
\begin{tabular}{cccccc}
\hline
\hline
ID &  Date & \nustar~Obs. ID & Exposure (s) & \swift~Obs. ID & Exposure (s)\\ \hline
\multicolumn{6}{c}{LMC~X--1} \\ \hline
A1 & 2014-08-08 & 30001039002 & 39971 & 00080763001 & 2715 \\
A2 & 2014-11-09 & 30001143002 & 47789 & 00080851001 & 2138 \\
A3 & 2016-05-11 & 30201029002 & 40223 &00081900001 & 1853 \\ \hline
\multicolumn{6}{c}{LMC~X--3}  \\ \hline
B1 & 2015-12-29 & 30101052002 & 27166 & 00081663001 & 1918 \\
B2 & 2019-02-18 & 30402035002 & 21973 & 00088777001 & 2168 \\
B3 & 2019-04-18 & 30402035004 & 30665 & 00088777002 & 2587 \\
B4 & 2019-05-25 & 30402035006 & 23106 & 00088777003 & 2033 \\
B5 & 2019-06-21 & 30402035008 & 31906 & 00088777004 & 1858 \\
\hline
\hline
\end{tabular}
\end{table*}

\section{Observation and Data Reduction}
\label{sec:obs}

LMC~X--1 was observed with \nustar at three epochs between 2014 and 2016, whereas LMC~X--3 was observed at five epochs between 2015 and 2019. Though, LMC~X--3 was observed again with \nustar twice in 2019, these observations were performed in calibration mode. Therefore, we did not use these observations in the present study. All eight epochs of \nustar observations (three for LMC~X--1 and five for LMC~X--3) were carried out simultaneously with \swift. The details of the observations are presented in Table~\ref{tab:log}.

\subsection{\nustar}
\label{sec:nustar}
\nustar is a hard X-ray focusing telescope, consisting of two identical modules: FPMA and FPMB \citep{Harrison2013}. The observations of the extra-galactic black hole candidates LMC~X--1 and LMC~X--3 were carried out with \nustar in SCIENCE observation mode. We first reprocessed the raw data using the \nustar Data Analysis Software ({\tt NuSTARDAS}, version 1.4.1). Cleaned event files were generated by using the standard filtering criteria in the {\tt nupipeline} task and the latest calibration data file (version 20200813) available in the {\it NuSTAR} calibration database (CALDB) \footnote{\url{http://heasarc.gsfc.nasa.gov/FTP/caldb/data/nustar/fpm/}}. We extracted source products by considering circular regions of radius 60 arcsec centred at the source coordinates. A circular region of 90 arcsec radius, located away from the source position, was chosen as the background region. We extracted the spectra and light curves using the {\tt nuproduct} task. The spectra were re-binned by using the {\tt grppha} task to achieve 20 counts per bin. As both the sources are weak in hard X-rays and data beyond 30 keV are dominated by background, we restricted our spectral fitting in $3-30$~keV range. 

\subsection{\swift}
\label{sec:swift}

\swift observed LMC~X--1 and LMC~X--3 simultaneously with \nustar in window-timing (WT) mode. The $0.5-8$~keV spectra were generated using the standard online tools provided by the UK {\it Swift} Science Data Centre \citep{Evans2009} \footnote{\url{http://www.swift.ac.uk/user_objects/}}.

\begin{figure*}
\centering
\includegraphics[width=8.5cm]{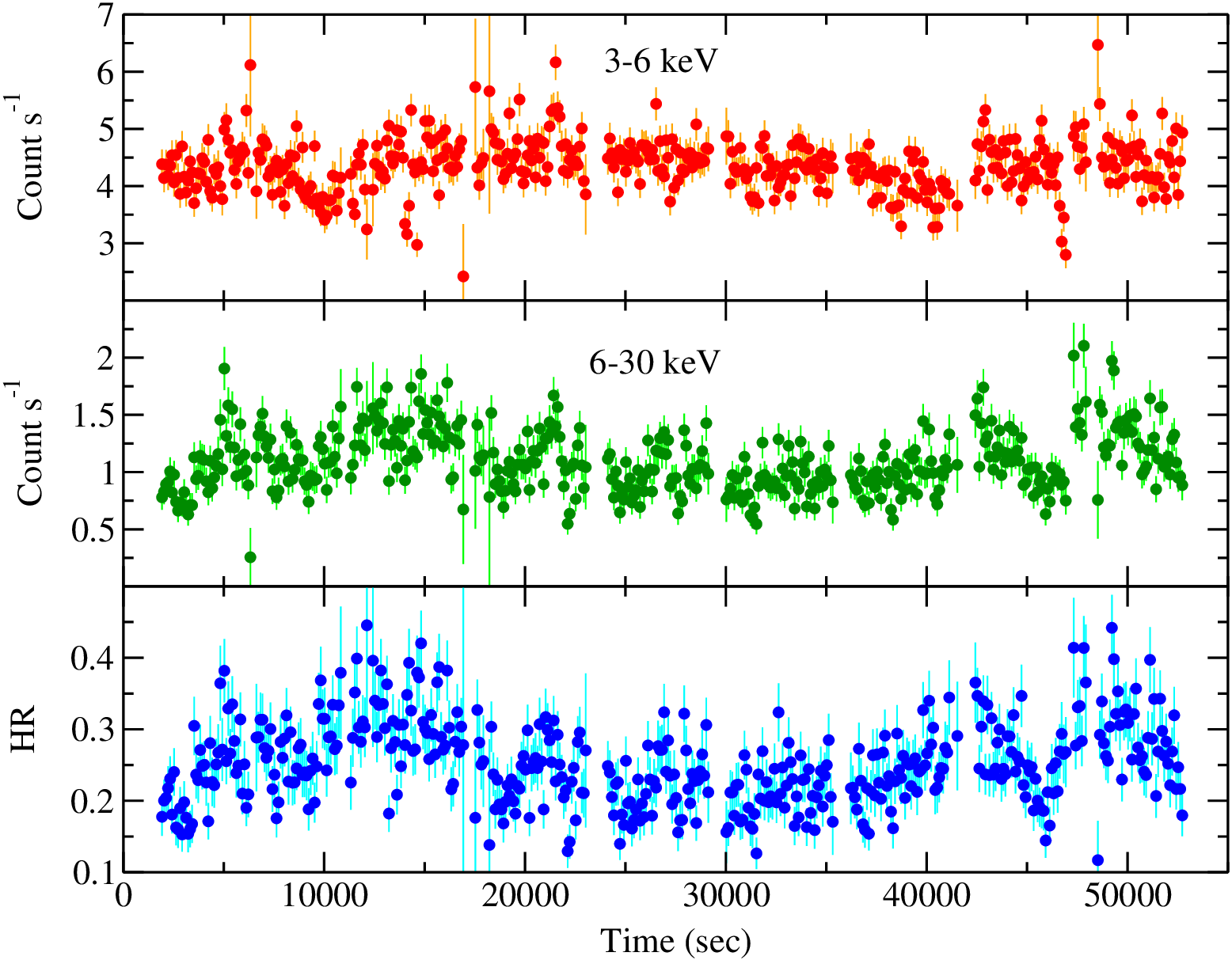}\hskip0.5cm
\includegraphics[width=8.5cm]{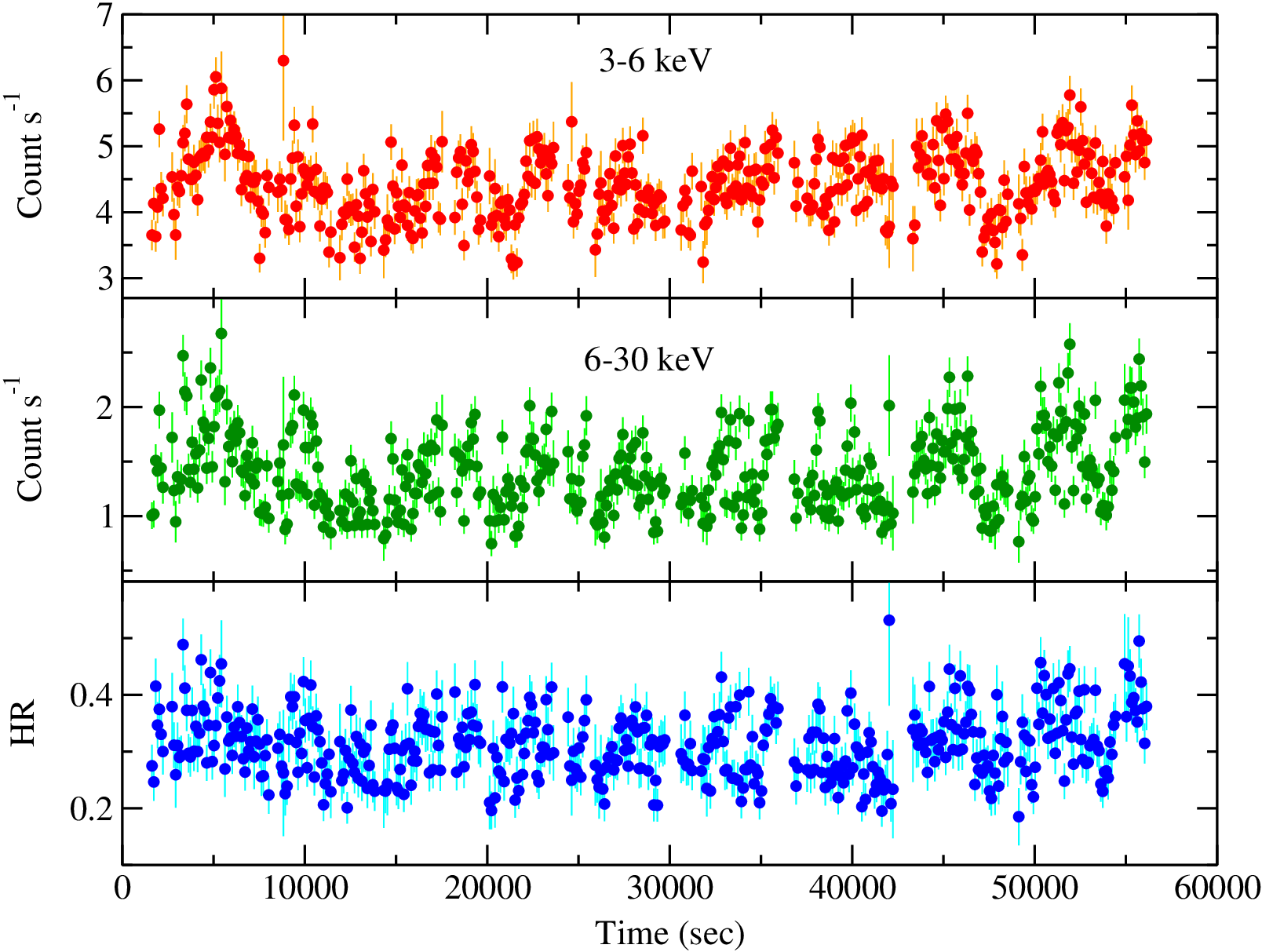}
\caption{Representative \nustar~ light curves in $3-6$~keV range (top panels), $6-30$~keV range (middle panels) and hardness-ratio (HR; bottom panels) are shown for LMC~X--1 for the observation A1 (left panels) and LMC~X--3 for the observation B1 (right panels) are shown. The HR is defined as the ratio between the count rates in $6-30$~keV and $3-6$~keV ranges.}
\label{fig:lc}
\end{figure*}

\section{Analysis \& Results}
\label{sec:res}

\subsection{Timing Properties}
\label{sec:timing-prop}

\begin{figure*}
\centering
\includegraphics[width=8.5cm]{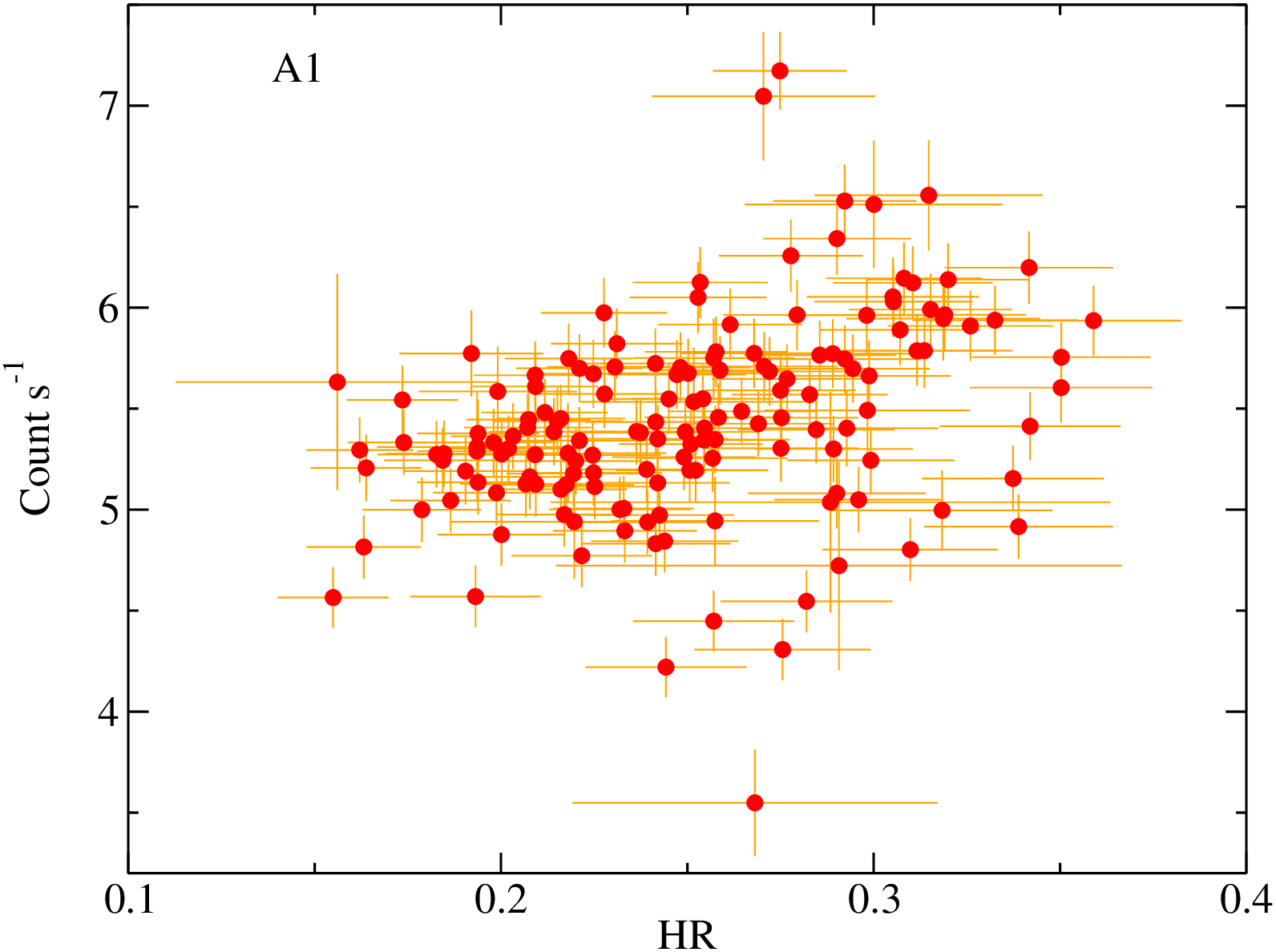}\hskip0.5cm
\includegraphics[width=8.5cm]{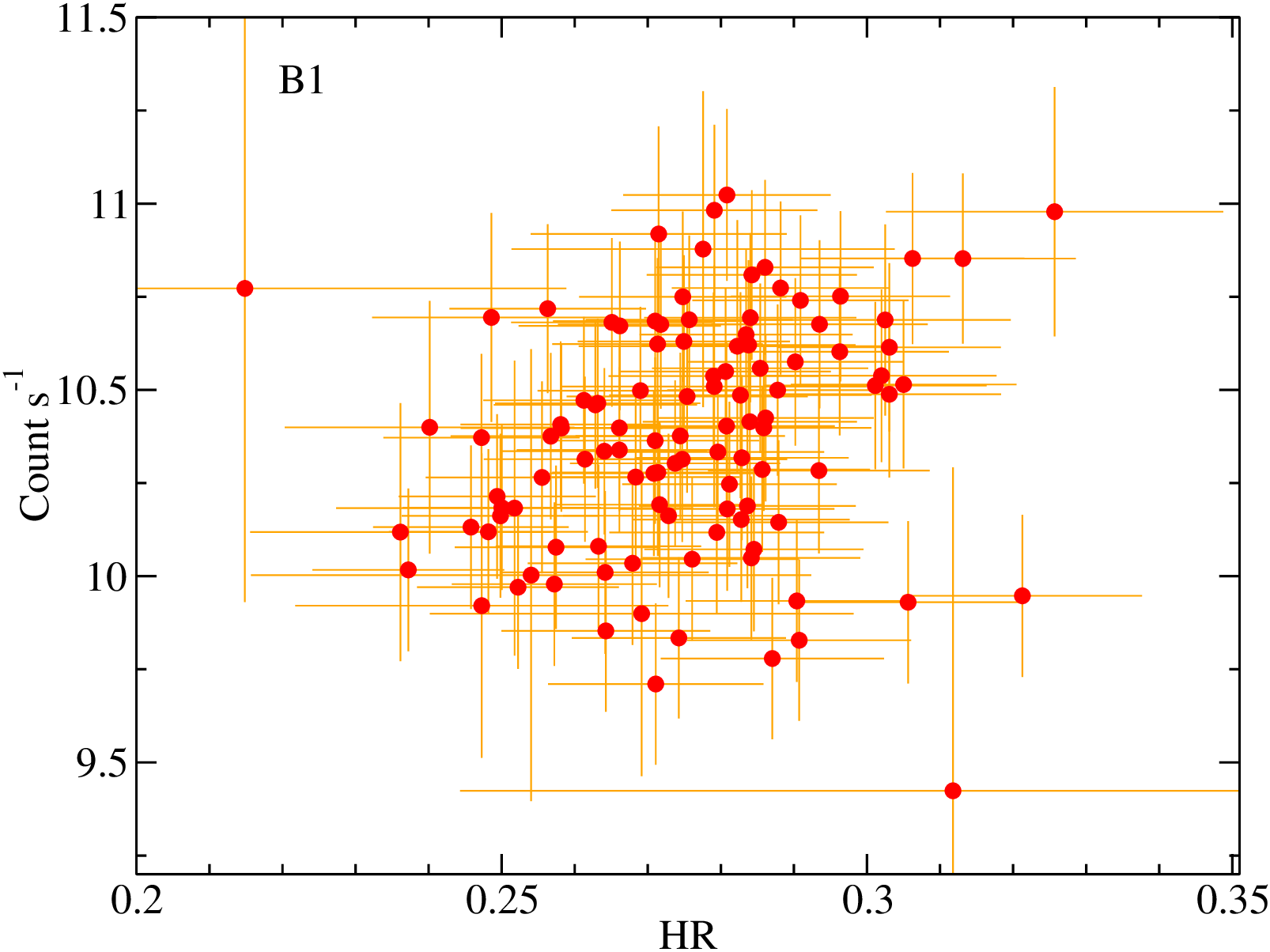}
\caption{Hardness-intensity diagrams (HID) are shown for LMC~X--1 (for observation A1) and for LMC~X--3 (for observation B1) in the left and right panels, respectively. The $3-30$~keV count rate is plotted as a function of hardness ratio (HR) in both the panels.}
\label{fig:hid}
\end{figure*}

We generated background-subtracted \nustar~ light curves in the energy ranges of 3 -- 30~keV, 3 -- 6~keV and 6 -- 30~keV for both LMC~X--1 and LMC~X--3 at a time resolution of 0.01~s. We calculated the hardness ratio (HR) by taking the ratio between the light curves in 6 -- 30 keV and 3 -- 6 keV ranges. A BHXB spectrum mainly consists of a multi-colour thermal disc blackbody component and a hard Comptonized emission. In general, the thermal emission dominates below 6 keV while the Comptonized emission dominates above it. Hence, we chose 3 -- 6 keV and 6 -- 30 keV ranges for the soft and hard X-ray bands, respectively, to study the variation of HR. The nature of the light curves (average count rate and variabilities) and hardness ratios are found to be comparable during all the epochs of observations of each source. The average count rate was observed to be higher in the 3 -- 6 keV energy range during all the epochs of observations for both sources. In Figures~\ref{fig:lc}a and ~\ref{fig:lc}b, we show the representative light curves in 3 -- 6 keV range, 6 -- 30  keV range and hardness ratio in the top, middle and bottom panels for LMC~X--1 (2014 August 8 \nustar observation: A1) and LMC~X--3 (2015 December 29 \nustar observation: B1), respectively. We observed that the HR was in the range of $\sim 0.1-0.4$ for both sources during all the observations. In general, a weak trend of increasing HR with count rate was observed in all observations of both sources. Representative hardness-intensity diagram (HID) for LMC~X--1 (A1) and LMC~X--3 (B1) are shown in the left and right panels of Figure~\ref{fig:hid}, respectively, where the $3-30$~keV count rate is plotted as a function of the HR.

We generated power density spectra (PDS) by applying the Fast Fourier Transformation (FFT) technique on the $3-30$~keV light curves using {\tt powspec} task of {\tt FTOOLS}. We divided the light curves into 8192 intervals and generated the Poisson noise subtracted PDS for each interval. The final PDS are generated by averaging all the PDS for each observation. We re-binned the PDS with a factor of 1.05 in a geometrical series manner. The final PDS are normalized to Miyamoto normalization where the fractional rms spectra are in the unit of $(rms/mean)^2 Hz^{-1}$ \citep{Miyamoto1991,vanderklis1997}. The 0.01~s light curves allowed the PDS up to Nyquist frequency of 50~Hz. However, in the case of both LMC~X--1 and LMC~X--3, the power diminishes above 10~Hz. We calculated fractional rms amplitude for each PDS of both the sources by integrating power in the range of $0.01-10$~Hz \citep{vanderklis2004}. The variability in the PDS was found to be weak for both the source during all epoch of observations with fractional rms amplitudes of $<2\%$.

\begin{figure*}
\centering
\includegraphics[width=8.5cm]{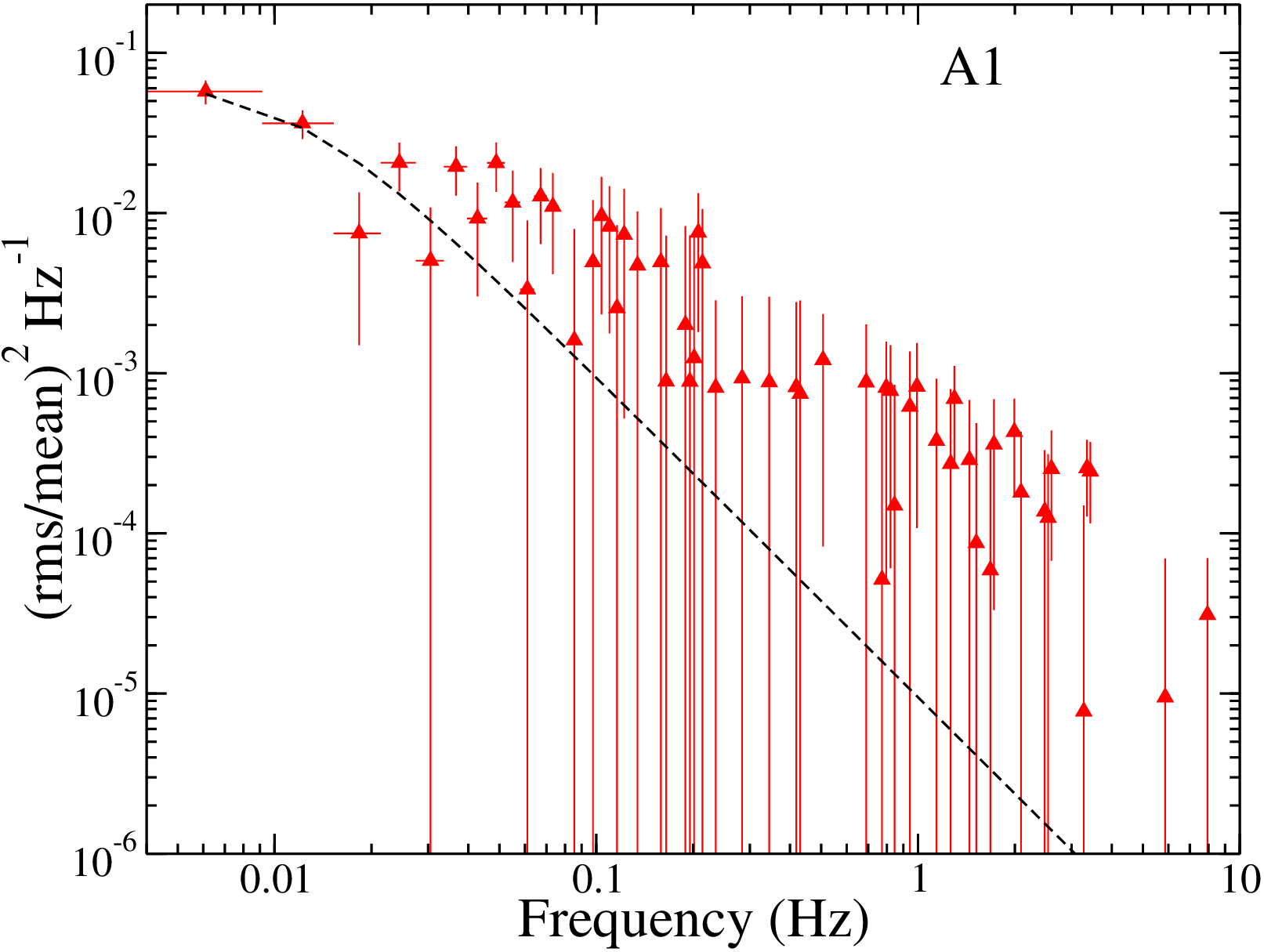}\hskip0.5cm
\includegraphics[width=8.5cm]{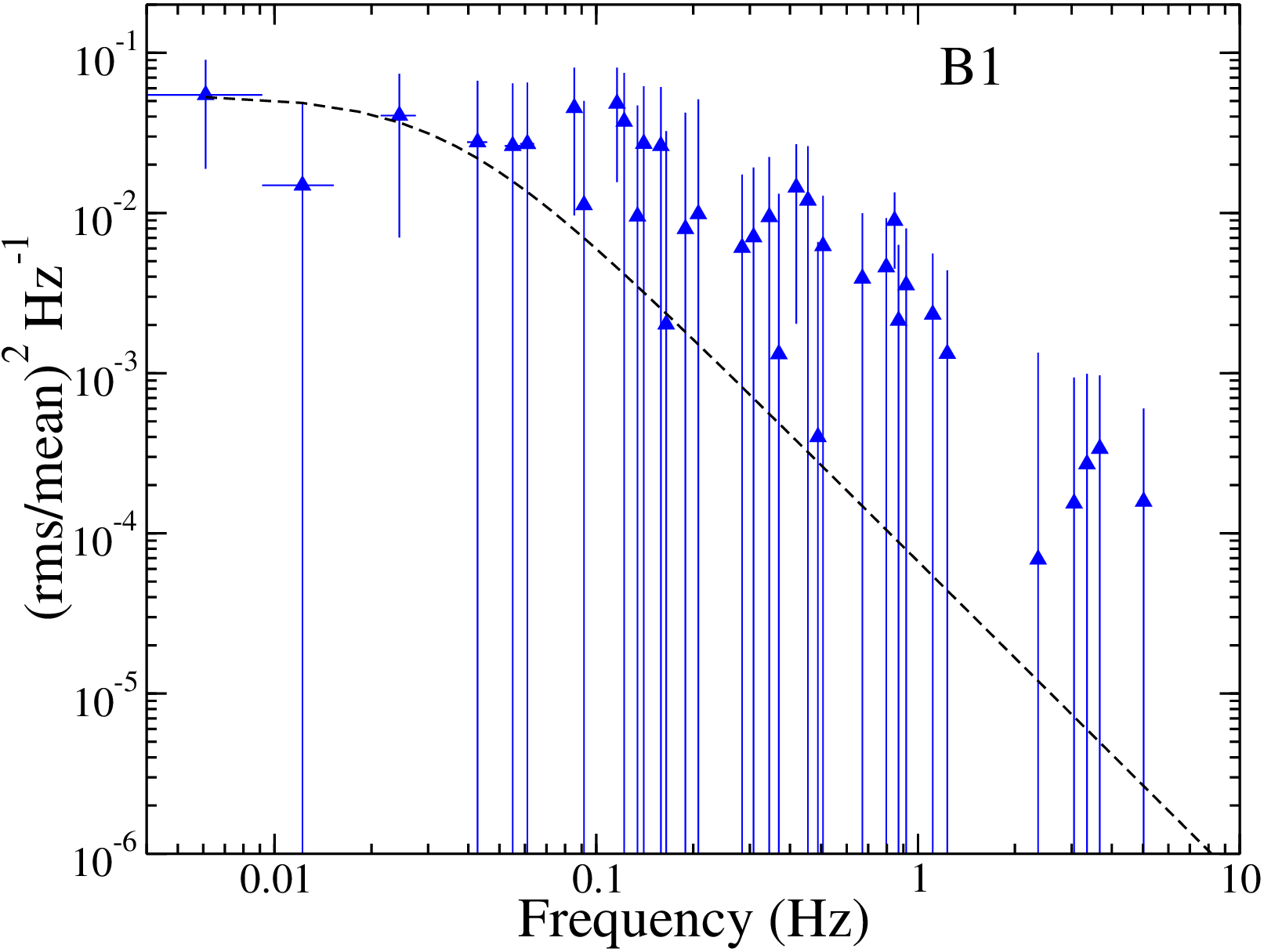}
\caption{The Poisson noise subtracted power density spectra for LMC~X--1 (for observation A1) and LMC~X--3 (for observation B1) are shown in the left and right panels, respectively. The black dashed curves represent the best-fitted Lorentzian function.}
\label{fig:a1-pds}
\end{figure*}

We did not find any evidence of quasi-periodic oscillation (QPO) or peaked noise in the PDS of both sources. Thus, we calculated the characteristic frequency ($\nu_{\rm C}$) of each PDS. The characteristic frequency is the frequency where the Lorentzian contributes maximum to the total rms. The characteristic frequency represents the broadband noise, similar to the QPO frequency, which expresses the peaked noise. We fitted each PDS with a zero-centered Lorentzian to calculate the characteristic frequency. The characteristic frequency is given by, $\nu_{\rm C} = \sqrt{\nu_{\rm 0} + (\Delta \nu /2)^2}$, where $\nu_{\rm 0}$ and $\Delta \nu$ are the centroid frequency and full width at half maximum of the Lorentzian, respectively. The characteristic frequency is found to be $\nu_{\rm C} \sim 0.02$~Hz for LMC~X--1 and $\nu_{\rm C}\sim 0.03$~Hz for LMC~X--3, respectively. Representative PDSs for LMC~X--1 (A1) and LMC~X--3 (B1) are shown in Figure~\ref{fig:a1-pds}. The dotted curve in each panel represents the best-fitted Lorentzian. 

\subsection{Spectral Analysis}
\label{sec:spec-analysis}
We carried out spectral analysis of the data in $0.5-30$~keV range, obtained from \swift/XRT ($0.5-8$~keV range) and \nustar ($3-30$~keV range) observations of LMC~X--1 and LMC~X--3. The spectral analysis was done using {\tt HEASARC}'s spectral analysis software package {\tt XSPEC} v12.10\footnote{\url{https://heasarc.gsfc.nasa.gov/docs/xanadu/xspec/}}\citep{Arnaud1996}. In our analysis, we used {\tt diskbb} \citep{Mitsuda1984} and {\tt kerrbb} \citep{Li2005} models for the thermal component, {\tt nthcomp} \citep{Zycki1999} and {\tt simpl} \citep{Steiner2009} models for the Comptonized component and {\tt relxill} for the reflection. We also used {\tt edge} for the absorption edge, {\tt Gaussian} line for the Fe K-line emission, and {\tt TBabs} for the interstellar absorption in the spectral fitting. We used {\it wilms} abundances \citep{Wilms2000} with cross-section of Verner \citep{Verner1996} for the absorption.

We started spectral analysis of LMC~X--1 with a model consisting of an absorbed thermal and Comptonized emission components. The thermal component was modelled with the {\tt diskbb}, while the Comptonized emission was modelled with {\tt nthcomp}. We refer to this as Model-1. In the case of LMC~X--1, we required a {\tt Gaussian} component to incorporate the Fe K$\alpha$ emission line in the spectrum. In the XRT spectrum of LMC~X--3, we observed an absorption edge at $\sim 1$~keV, which was approximated with {\tt edge} component in the {\tt XSPEC} package. The Model-1 reads as {\tt constant*tbabs*(diskbb+nthcomp+ga)} and {\tt constant*tbabs*edge*(diskbb+nthcomp)} for LMC~X--1 and LMC~X--3, respectively. The {\tt constant} factor is included as the cross-normalization factor between the \swift/XRT and \nustar. The cross-normalization factor for \swift/XRT was fixed at 1 while it was allowed to vary for \nustar. The cross-normalization factor for \nustar was found in the range of $0.95-0.98$ for both sources. In our fitting, we linked seed photon temperature ($kT_{\rm bb}$ of {\tt nthcomp}) with the inner disc temperature ($T_{\rm in}$). During the fitting, we could not constrain the hot electron plasma temperature ($kT_{\rm e}$) and thus fixed it at 1000~keV. The parameters obtained from fitting the data with Model-1 are given in Tables~\ref{tab:lmcx1} \& ~\ref{tab:lmcx3} for LMC~X--1 and LMC~X--3, respectively. Though the spectral fitting of data in $0.5-30$~keV range with Model-1 was acceptable, we explored other models to fit the data of both sources. 

Next, we used a relativistic thin accretion disc model {\tt kerrbb} for the thermal emission. For the Comptonized emission component, we used convolution model {\tt simpl} that describes the Comptonized emission by up-scattering seed photons from the thermal component \citep{Steiner2009}. We refer to this model as Model-2 which reads in {\tt XSPEC} as {\tt constant*tbabs*(simpl*kerrbb+ga)} for LMC~X--1, and {\tt constant*tbabs*edge*(simpl*kerrbb)} for LMC~X--3, respectively. While fitting the data with Model-2, we kept the disc inclination angle and distance fixed at the corresponding reported values for both the sources \citep{Orosz2009,Orosz2014}. It is to be noted that we assumed the disc inclination is the same as the inclination of the binary system, i.e. the disc is not warped. The assumption is made following the previous work on LMC~X--1 and LMC~X--3 \citep{Gou2009,Steiner2014,Bhuvana2021}. We also assumed zero torque at the inner edge. As recommended, we fixed the model normalization to 1 and allowed mass, spin, accretion rate and spectral hardening factor to vary freely. The parameters obtained from fitting data with model are given in Tables~\ref{tab:lmcx1} \& ~\ref{tab:lmcx3} for LMC~X--1 and LMC~X--3, respectively.

Since the reflection hump is likely to be observed in $\sim15-30$~keV energy range, we used relativistic reflection model {\tt relxill} \citep{Garcia2013,Dauser2016} to probe reflection component in the data. The {\tt relxill} describes the reflection along with the Comptonized emission. The relative reflection fraction ($R_{\rm refl}$) measures the reflection strength and is defined as the ratio of Comptonized emission to the distant observer to the reflected emission. {\tt Relxill} assumes a broken power-law emission profile, where $E(r) \sim R^{-q_1}$ for $R < R_{\rm br}$ and $E(r) \sim R^{-q_2}$ for $R > R_{\rm br}$. Here $E(r)$, $R_{\rm br}$ and $q$ are the emissivity, break radius and emissivity index, respectively. This model is termed as Model-3 which reads in {\tt XSPEC} as {\tt constant * tbabs * (kerrbb + relxill)} for LMC~X--1 and {\tt constant * tbabs * edge*(kerrbb + relxill)} for LMC~X--3.  We linked the spin parameters and inclination angle between the {\tt kerrbb} and {\tt relxill} models. During our analysis, the outer edge of disc was fixed at the typical value of $R_{\rm out}=1000~R_{\rm g}$. The inner edge ($R_{\rm in}$) was fixed to inner most stable circular orbit or ISCO ($R_{\rm ISCO}$) assuming the disc extends to the ISCO. We froze the emissivity index, $q_2=3$ for $R>R_{\rm br}$, during our analysis. We also fixed the cut-off energy at $E_{\rm cut}=400$~keV as we could not constrain it while fitting the $0.5-30$~keV spectra. The iron abundance was kept frozen at the Solar value, i.e. $A_{\rm Fe}=1~A_{\sun}$. The inclination angle was frozen at the reported values of $36.38^{\circ{}}$ \citep{Orosz2009} and $69.24^{\circ{}}$ \citep{Orosz2014} for LMC~X--1 and LMC~X--3, respectively. The parameters obtained from the spectral fitting are quoted in Tables~\ref{tab:lmcx1} \& ~\ref{tab:lmcx3} for LMC~X--1 and LMC~X--3, respectively.

\subsection{Spectral Properties}
\label{sec:spec-prop}

\subsubsection{LMC~X--1}
\label{sec:lmc-x1}
We obtained a good fit while fitting the spectrum of LMC~X--1 at all the epochs of the observations with Model-1. The hydrogen column density was found to be $N_{\rm H} \sim 9.1-9.6~ \times~ 10^{21}$ \pcm  which is marginally less compared to the previous report of $N_{\rm H} \sim 1-1.3 ~\times 10^{22}$ \pcm \citep{Hanke2010}. The spectral analysis revealed that the inner disc temperature varied in the range of $T_{\rm in} \sim 0.85-1$~keV, with the photon index $\Gamma \sim 2.5-3.5$. We detected a broad Fe K$\alpha$ emission line at $\sim 6.4$~keV with line width, $\sigma \sim 1$~keV in all three epoch of observations. The equivalent width (EW) for Fe K$\alpha$ emission line were obtained to be $247^{+6}_{-4}$~eV, $262^{+6}_{-5}$~eV and $227^{+4}_{-5}$~eV for A1, A2 and A3, respectively. Detection of broad Fe K-line in the spectra indicated that the Fe line is originated in the accretion disc and the broadening is due to the gravitational redshift and Doppler effect. The source was found in the thermal dominated HSS during all three observations, with the fraction of the thermal emission $f_{\rm disc} = F_{\rm disc}/F_{\rm tot}>87\%$. Here $F_{\rm disc}$ and $F_{\rm tot}$ are the thermal disc flux and total flux, respectively.

While fitting the data with Model-2, we obtained accretion rate ($\dot{\rm M}$), mass ($M_{\rm BH}$) and spin parameter ($a^*$) of the black hole. The best-fitted mass is found to be in the range of $M_{\rm BH} \sim 10.7-11.1$ $M_{\odot}$, which is consistent with dynamical measurement of mass, $M_{\rm BH} = 10.91 \pm 1.41~M_{\sun}$ \citep{Orosz2009}. We found that the BH in LMC~X--1 is a high spinning BH with, $a^* \sim 0.93-0.95$. The spectral hardening factor and mass accretion rate are found to be, $\kappa \sim 1.53-1.58$ and $\dot{\rm M} \sim 1-1.3 \times 10^{18}$ \gps, respectively. The photon index was found in the range of $\Gamma \sim 2.7-3.5$ with the scattering fraction $f_{\rm Scat} \sim 0.1-0.13$.

We found the evidence of a very weak reflection with $R_{\rm refl}<0.13$ in all three epoch of observations while fitting with Model-3. The emissivity index for $R<R_{\rm br}$ and the break radius were obtained to be $q_1 \sim 4-5$ and $R_{\rm br} \sim 10~R_{\rm g}$, respectively. The disc was found to be moderately ionized with $\xi \sim 10^{2.6-3}$ \ecps.

\begin{figure*}
\centering
\includegraphics[width=8.5cm]{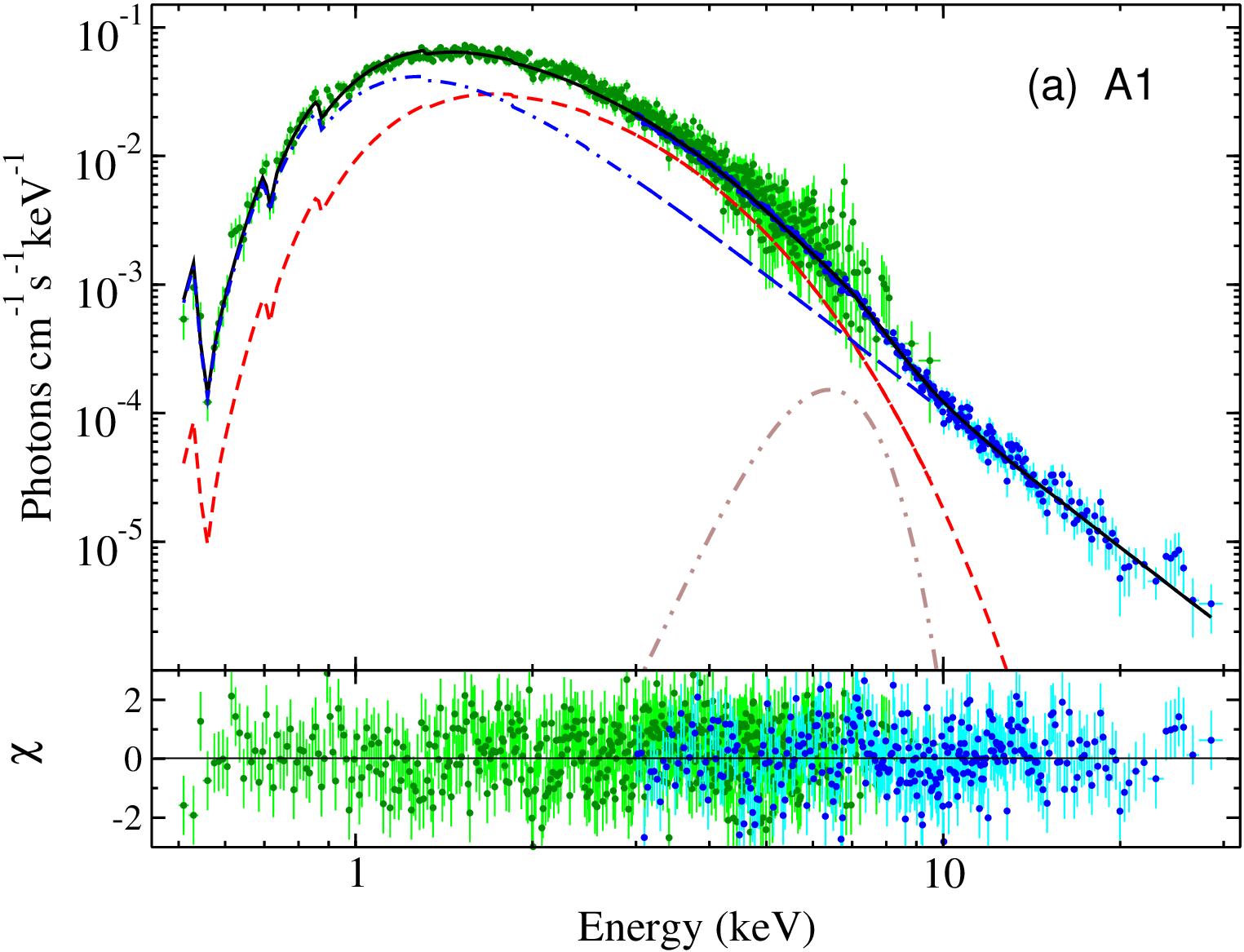}\hskip0.5cm
\includegraphics[width=8.5cm]{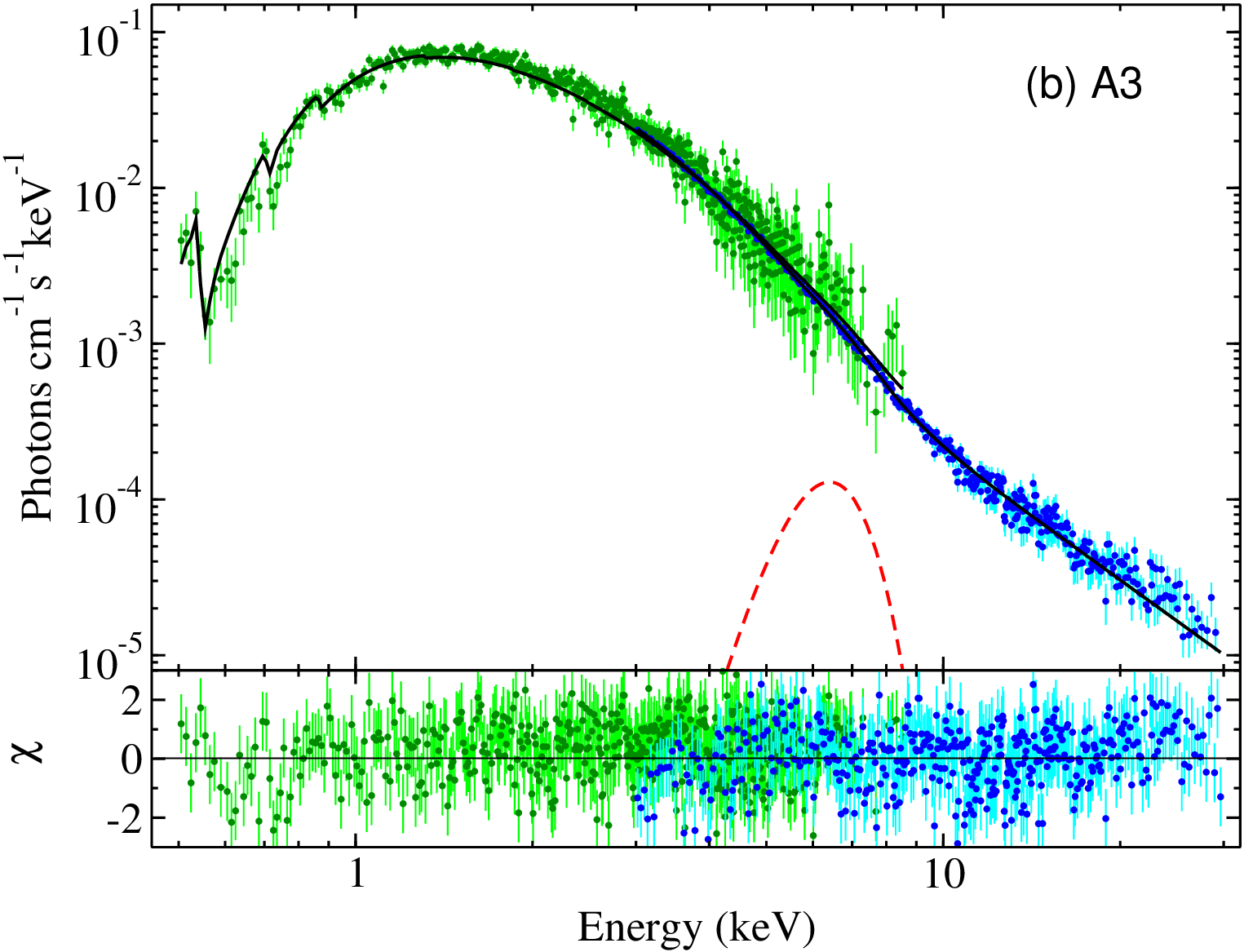}
\caption{Left panel: Best fitted spectrum of LMC~X--1 along with Model-2 is shown for 2014 August 8 \nustar and \swift observations (A1). The black solid, red dashed, blue dot-dashed and brown dot-dot-dashed lines represent the total emission, thermal component, Comptonized component and Fe K$\alpha$ line emission, respectively. Right panel: Best fitted spectrum of LMC~X--1 along with Model-2 is shown for 2016 May 11 \nustar and \swift observations (A3). The black solid and red dashed lines represent the total emission and Fe K$\alpha$ line emission, respectively. Corresponding residuals are shown in the bottom panel of each spectrum.}
\label{fig:a-spec}
\end{figure*}

\begin{table*}
\caption{Best-fitted spectral parameters of LMC~X--1.}
\label{tab:lmcx1}
\begin{tabular}{cccccccccc}
\hline
\multicolumn{10}{c}{Model - 1 : TBABS*(DISKBB+NTHCOMP+GA)} \\ \hline
ID & $N_{\rm H}$& $T_{\rm in}$ & $N_{\rm DBB}$ & $\Gamma$ & $N_{\rm nthcomp}$  & LN & $L_{\rm X}$ & $f_{\rm disc}$ &$\chi^2$/dof \\ 
 & ($10^{22}$ \pcm) & (keV) &  & & ($10^{-2}$) & ($10^{-4}$) & ($10^{38}$ \eps) &  (\%) &   \\
\hline
A1 & $0.96^{+0.02}_{-0.02}$ & $0.85^{+0.03}_{-0.01}$ & $59.1^{+3.6}_{-3.7}$ & $3.55^{+0.20}_{-0.18}$ & $2.70^{+0.77}_{-0.40}$ & $5.43^{+0.10}_{-0.07}$ & $1.38^{+0.03}_{-0.04}$ & $ >87$ & 1175/1004 \\
A2 & $0.91^{+0.03}_{-0.04}$ & $0.96^{+0.01}_{-0.02}$ & $62.9^{+1.5}_{-1.4}$ & $2.78^{+0.06}_{-0.05}$ & $1.89^{+0.21}_{-0.19}$ & $5.57^{+0.12}_{-0.09}$ & $1.40^{+0.02}_{-0.02}$&$ >89$ & 1304/1111 \\
A3 & $0.91^{+0.02}_{-0.02}$ & $0.83^{+0.01}_{-0.01}$ & $67.9^{+3.6}_{-3.4}$ & $2.57^{+0.08}_{-0.08}$ & $1.64^{+0.20}_{-0.23}$ & $7.58^{+0.18}_{-0.21}$ & $1.52^{+0.03}_{-0.04}$&$ >92$ & 1213/1077 \\
\hline
\multicolumn{10}{c}{Model - 2 : TBABS*(KERRBB*SIMPL+GA)}  \\ \hline
ID & \nh & $a^*$ & $M_{\rm BH}$ & $\dot{\rm M}$ & $\kappa$ & $\Gamma$ & $f_{\rm Scat}$ & LN & $\chi^2$/dof \\
& ($10^{22}$ \pcm)& & ($M_{\odot}$) &($10^{18}$ gm s$^{-1}$)& & & & ($10^{-4}$) &  \\
\hline
A1& $0.97^{+0.02}_{-0.03}$&$0.95^{+0.01}_{-0.01}$& $11.0^{+0.4}_{-0.3}$&$ 1.06^{+0.04}_{-0.08}$&$ 1.53^{+0.02}_{-0.01}$&$ 3.48^{+0.10}_{-0.18}$&$ 0.13^{+0.01}_{-0.02}$&$ 1.38^{+0.03}_{-0.04}$& 1055/1003\\
A2& $0.89^{+0.05}_{-0.03}$&$0.93^{+0.05}_{-0.03}$& $10.7^{+0.6}_{-0.4}$ &$ 1.18^{+0.03}_{-0.05}$&$ 1.58^{+0.02}_{-0.02}$&$ 2.71^{+0.08}_{-0.12}$&$ 0.11^{+0.01}_{-0.01}$& $ 1.40^{+0.02}_{-0.02}$& 1727/1110\\
A3& $0.89^{+0.04}_{-0.05}$&$0.95^{+0.03}_{-0.02}$& $11.1^{+0.4}_{-0.5}$&$ 1.31^{+0.03}_{-0.04}$&$ 1.55^{+0.02}_{-0.03}$&$ 2.69^{+0.06}_{-0.13}$&$ 0.10^{+0.02}_{-0.01}$& $1.52^{+0.03}_{-0.04}$& 1173/1078\\ 
\hline
\multicolumn{10}{c}{Model - 3 : TBABS*(KERRBB+RELXILL)}   \\ \hline
ID & $N_{\rm H}$ &$a^*$ & $\dot{\rm M}$ & $\kappa$ &$\log {\rm \xi}$ & $\Gamma$ & $R_{\rm refl}$ &$N_{\rm rel}$ & $\chi^2$/dof \\
 & ($10^{22}$ \pcm)& & ($10^{18}$ gm s$^{-1}$) & & & & & ($10^{-3}$) & \\
\hline
A1& $0.96^{+0.03}_{-0.02}$&$0.94^{+0.03}_{-0.03}$&$1.08^{+0.03}_{-0.04}$&$1.55^{+0.02}_{-0.02}$& $2.99^{+0.08}_{-0.10}$ &$3.24^{+0.10}_{-0.14}$&$<0.13$ &$8.15^{+0.64}_{-0.83}$ &1122/1001\\
A2& $0.91^{+0.02}_{-0.03}$  &$0.93^{+0.03}_{-0.06}$&$1.19^{+0.04}_{-0.05}$&$1.56^{+0.03}_{-0.03}$ &$2.85^{+0.08}_{-0.11}$ &$2.71^{+0.12}_{-0.09}$&$<0.08$ &$7.03^{+0.88}_{-1.05}$ &1388/1108\\
A3& $0.90^{+0.03}_{-0.02}$ & $0.95^{+0.02}_{-0.03}$&$1.33^{+0.04}_{-0.04}$&$1.55^{+0.02}_{-0.03}$&$2.65^{+0.11}_{-0.09}$ &$2.63^{+0.12}_{-0.16}$&$<0.12$ &$4.97^{+0.92}_{-0.64}$ &1178/1075\\
\hline
\end{tabular}
\leftline{The errors are quoted for 90\% confidence level (1.6 $\sigma$).}
\leftline{\nh is the hydrogen column density in $10^{22}$ \pcm. $T_{\rm in}$ and $N_{\rm DBB}$ are the inner disc temperature (in keV) and normalization of {\tt DISKBB} model, respectively.}
\leftline{$\Gamma$, $N_{\rm nthcomp}$ are the photon index and normalization of {\tt NTHCOMP} model, respectively. LN is the normalization (in $10^{-4}$ \phc) of the {\tt Gaussian} line.}
\leftline{$L_{\rm X}$ is the broadband X-ray luminosity (in $10^{38}$ \eps) in the energy range of $0.1-100$~keV. $f_{\rm disc}$ is the thermal fraction and defined as
$f_{\rm disc} = F_{\rm disc}/F_{\rm tot}$, }
\leftline{where $F_{\rm disc}$ and $F_{\rm tot}$ are disc flux and total flux, respectively.}
\leftline{$a^*$, $M_{\rm BH}$, $\dot{\rm M}$, $\kappa$ are the dimensionless spin parameter, mass of the black hole (in $M_{\odot}$), mass accretion rate (in $10^{18}$ \gps), and spectral hardening }
\leftline{factor of {\tt KERRBB} model respectively.}
\leftline{$\Gamma$ and $f_{\rm Scat}$ are the photon index and scattering fraction of {\tt SIMPL} model, respectively.}
\leftline{${\rm \xi}$, $R_{\rm refl}$ and $N_{\rm rel}$ are the ionization parameter (in \ecps), reflection fraction and normalization (in $10^{-3}$ \phc) of {\tt RELXILL} model,}
\leftline{respectively.}
\end{table*}

\begin{figure*}
\centering
\includegraphics[width=8.5cm]{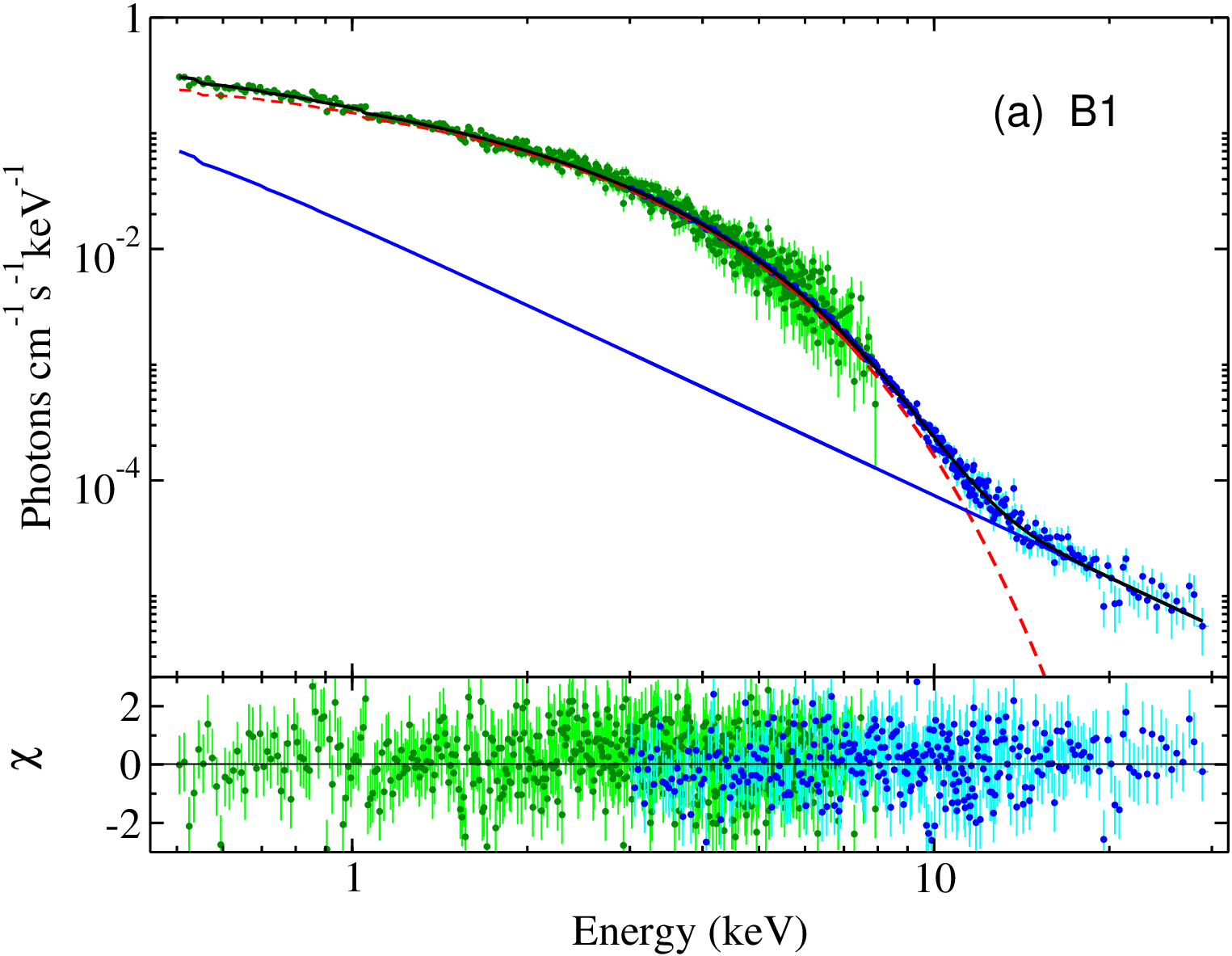}\hskip0.5cm
\includegraphics[width=8.5cm]{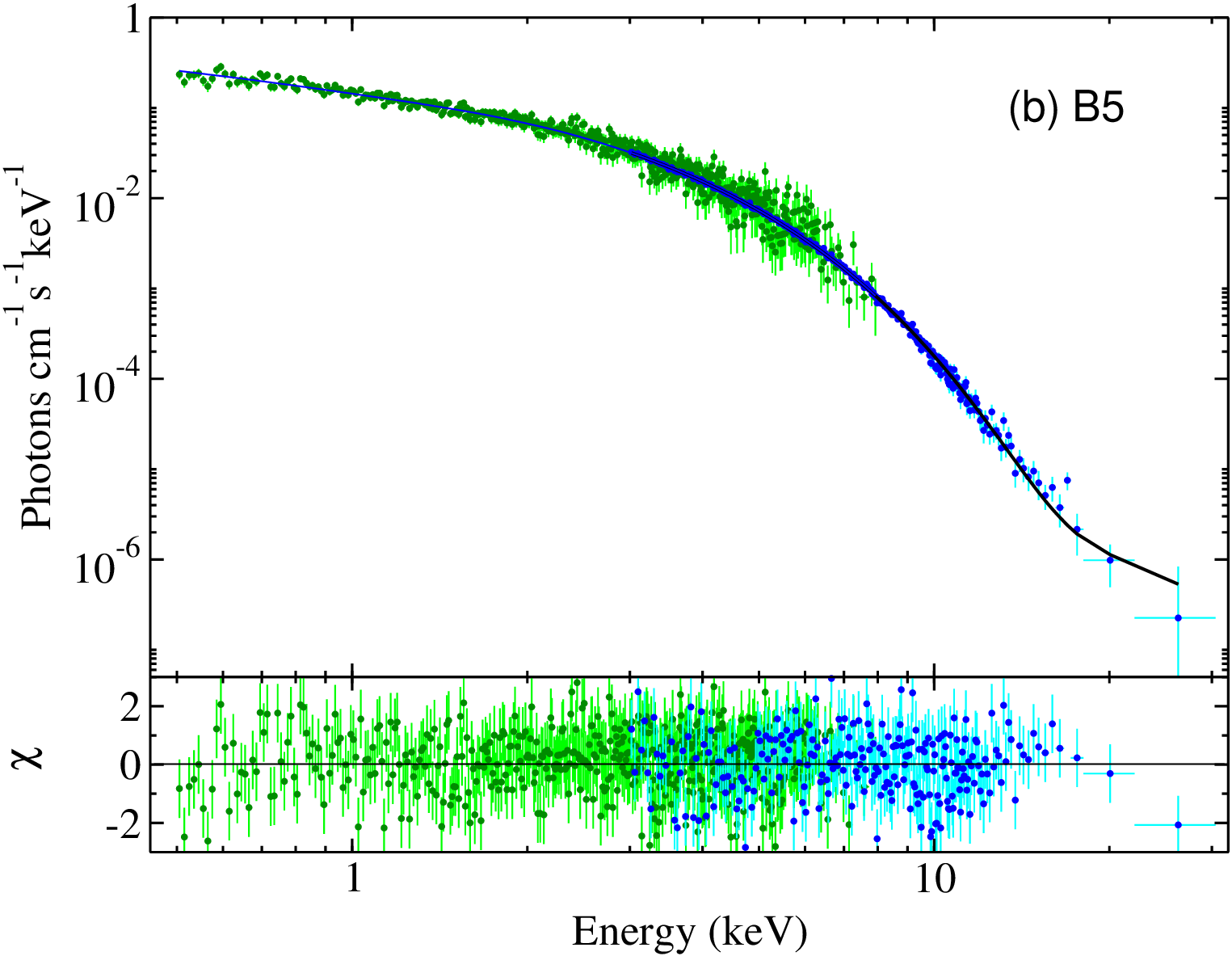}
\caption{Left panel: Best fitted spectrum of LMC~X--3 along with Model-2 is shown for 2015 December 29 \nustar and \swift observations (B1). The black solid, red dashed and blue solid lines represent the total emission, thermal component and Comptonized emission, respectively. Right panel: Best fitted spectrum of LMC~X--3 along with Model-3 is shown for 2019 June 21 \nustar and \swift observations (B5). The black solid line represents the total emission. Corresponding residuals are shown in the bottom panel of each spectrum.}
\label{fig:b-spec}
\end{figure*}

\subsubsection{LMC~X--3}
\label{sec:lmc-x3}
We started the spectral analysis of LMC~X--3 with Model-1 as describe in Section~\ref{sec:spec-analysis}. During all the observations, we obtained low hydrogen column density with $N_{\rm H} \sim 5-6 \times 10^{20}$ \pcm. An absorption edge was observed at $\sim 1$~keV in the spectrum of LMC~X--3 during observations B1, B2 and B3. The inner disc temperature was obtained to be in the range of, $T_{\rm in} \sim 1-1.2$~keV with the disc normalization, $N_{\rm DBB} \sim 24-30$. The photon index varied in the range of $\Gamma \sim 2.4-3.9$. Our spectral analysis revealed that all the five observations with \nustar and \swift observatories were carried out when the source was in the thermal dominated HSS. We observed that the disc contribution to the flux was very high with $f_{\rm disc}>78\%$. The spectral hardening factor was found to be $\kappa \sim 1.7$. The mass accretion rate was estimated to be in the range of $\dot{\rm M} \sim 4.05-9.24 \times 10^{18}$ \gps and was highest during the observation B4. The $0.1-100$~keV broadband X-ray luminosity was found to vary in the range of $L_{\rm X}=1.46-4.06 \times 10^{38}$ \eps. No evidence of reflection was seen in the spectra with $R_{\rm refl}<0.06$. The spin of the BH is determined to be in the range of $a^* \sim 0.19-0.29$, indicating a slowly spinning BH in the binary system. Our findings are consistent with the earlier reported results \citep{Steiner2010,Bhuvana2021}. The estimated mass from the spectral fitting is in the range of $M_{\rm BH} \sim 6.7-7.2$ $M_{\sun}$ which is consistent with the dynamical measurement \citep{Orosz2014}.

We did not detect any evidence of Fe K$\alpha$ emission line in the spectra of LMC~X--3. Thus, we calculated the upper limit on the Fe K-line emission. To calculate the upper limit on the equivalent width (EW), we fitted the spectra with Model-1 and Model-2 by adding a narrow Gaussian line. We kept the line energy frozen at 6.2~keV, 6.4~keV, 6.5~keV and 6.7~keV and line width at $0.01$~keV, $0.05$~keV, $0.1$~keV and $0.2$~keV, and calculated the EW. We found that if an iron line was present, it is most likely to be located at 6.5~keV. Nonetheless, we found the upper limit on the EW is $<10$~eV during all observations.

During B1, B2 and B3, an absorption edge at $\sim$1~keV was observed in the spectra of LMC~X--3. The absorption edge is presumably due to the absorption due to the ISM. Previously, the Oxygen absorption edge at $\sim$0.6~keV was detected in the spectrum of LMC~X--3 using {\it XMM-Newton} observation \citep{Page2003}. However, due to inferior spectral resolution of \swift/XRT, we could not identify the absorption edge.

\begin{table*}
\caption{Best-fitted spectral parameters of LMC~X--3.}
\label{tab:lmcx3}
\begin{tabular}{cccccccccc}
\hline
\multicolumn{10}{c}{Model - 1 : TBABS*EDGE*(DISKBB+NTHCOMP)} \\ \hline
ID & $N_{\rm H}$& $E_{\rm edge}$ &$T_{\rm in}$ & $N_{\rm DBB}$ & $\Gamma$ & $N_{\rm nthcomp}$ & $L_{\rm X}$ & $f_{\rm disc}$ &  $\chi^2$/dof \\ 
 & ($10^{22}$ \pcm) & (keV) & (keV) &  & & ($10^{-2}$) & ($10^{38}$ \eps) & (\%) &   \\
\hline
B1& $0.06^{+0.02}_{-0.02}$ &$ 1.08^{+0.02}_{-0.04}$ &$1.15^{+0.03}_{-0.03}$& $27.0^{+1.2}_{-0.7}$& $2.36^{+0.05}_{-0.15}$& $0.39^{+0.02}_{-0.01}$&$2.82\pm0.23$& $>94$ & 1065/1006 \\ 
B2& $0.05^{+0.02}_{-0.01}$ &$ 1.13^{+0.04}_{-0.07}$ &$0.99^{+0.03}_{-0.02}$& $28.4^{+0.9}_{-1.5}$& $2.62^{+0.09}_{-0.16}$& $1.67^{+0.03}_{-0.08}$&$1.87\pm0.19$&$>78$&1297/1048 \\
B3& $0.06^{+0.01}_{-0.01}$ &$ 0.98^{+0.05}_{-0.03}$ &$1.23^{+0.03}_{-0.05}$& $25.4^{+1.3}_{-1.7}$& $3.05^{+0.38}_{-0.11}$& $2.04^{+0.13}_{-0.07}$&$1.46\pm0.21$&$>89$&1134/995  \\
B4& $0.06^{+0.01}_{-0.01}$ &  --                    &$1.24^{+0.03}_{-0.04}$& $30.0^{+1.1}_{-1.9}$& $3.93^{+0.18}_{-0.29}$& $1.31^{+0.14}_{-0.07}$&$4.06\pm0.35$&$>94$&1151/1009 \\
B5& $0.05^{+0.01}_{-0.02}$ &  --                    &$1.16^{+0.05}_{-0.03}$& $24.0^{+0.8}_{-1.3}$& $3.54^{+0.25}_{-0.32}$& $>0.95 $               &$1.95\pm0.24$&$>96$& 971/874  \\
\hline
\multicolumn{10}{c}{Model - 2 : TBABS*EDGE*(KERRBB*SIMPL)}  \\ \hline
ID & \nh & $E_{\rm edge}$ & $a^*$ & $M_{\rm BH}$ &$\dot{\rm M}$ & $\kappa$ & $\Gamma$ & $f_{\rm Scat}$ & $\chi^2$/dof \\
& ($10^{22}$ \pcm)&(keV) & &($M_{\odot}$) & ($10^{18}$ gm s$^{-1}$)& & &  \\
\hline
B1& $0.04^{+0.01}_{-0.02}$&$ 1.04^{+0.04}_{-0.05}$&$ 0.19^{+0.05}_{-0.06}$& $7.1^{+1.2}_{-0.8}$ &$ 7.37^{+0.22}_{-0.32}$&$ 1.65^{+0.04}_{-0.03}$&$ 2.16^{+0.13}_{-0.35}$&$ <0.05$& 1035/997 \\
B2& $0.04^{+0.03}_{-0.01}$&$ 1.09^{+0.05}_{-0.08}$&$ 0.26^{+0.02}_{-0.06}$& $6.7^{+0.7}_{-1.0}$ &$ 4.70^{+0.45}_{-0.34}$&$ 1.67^{+0.04}_{-0.02}$&$ 2.51^{+0.22}_{-0.18}$&$ 0.10^{+0.02}_{-0.04}$& 1254/1046\\
B3& $0.07^{+0.02}_{-0.05}$&$ 0.99^{+0.07}_{-0.02}$&$ 0.29^{+0.05}_{-0.05}$& $6.9^{+0.5}_{-1.2}$ &$ 4.05^{+0.15}_{-0.31}$&$ 1.71^{+0.03}_{-0.06}$&$ 2.81^{+0.28}_{-0.39}$&$ 0.04^{+0.01}_{-0.02}$& 1196/992 \\
B4& $0.08^{+0.02}_{-0.05}$&  --                   &$0.20^{+0.07}_{-0.04}$& $6.8^{+2.1}_{-2.8}$ &$ 9.24^{+0.14}_{-0.47}$&$ 1.70^{+0.02}_{-0.04}$&$ 3.39^{+0.18}_{-0.45}$&$ <0.02$& 1122/1008\\
B5& $0.06^{+0.03}_{-0.04}$&  --             &$ 0.28^{+0.03}_{-0.08}$& $7.2^{+1.3}_{-0.8}$ &$ 5.92^{+0.45}_{-0.21}$&$ 1.67^{+0.05}_{-0.06}$&$ 3.09^{+0.14}_{-0.34}$&$ <0.03$& 954/872  \\
\hline
\multicolumn{10}{c}{Model - 3 : TBABS*EDGE*(KERRBB+RELXILL)}   \\ \hline
ID & $N_{\rm H}$ & $a^*$ & $\dot{\rm M}$ & ${\rm \kappa}$ & $\log {\rm \xi}$ & $\Gamma$ & $R_{\rm refl}$ &$N_{\rm rel}$ & $\chi^2$/dof \\
 & ($10^{22}$ \pcm) & & ($10^{18}$ gm s$^{-1}$) & & & & & ($10^{-3}$) & \\
\hline
B1&$0.04^{+0.01}_{-0.02}$ &$0.21^{+0.05}_{-0.04} $&$7.36^{+0.97}_{-1.02}$&$1.69^{+0.02}_{-0.02} $ &$2.73^{+0.02}_{-0.02} $ &$1.82^{+0.08}_{-0.06} $ &$<0.11 $ &$5.82^{+0.02}_{-0.03} $ &1027/993\\
B2&$0.05^{+0.02}_{-0.01}$&$0.25^{+0.06}_{-0.07} $&$4.56^{+0.87}_{-0.65}$&$1.69^{+0.01}_{-0.01} $  &$3^{\rm f}$&$2.35^{+0.21}_{-0.17} $ &$<0.06$ &$5.21^{+0.03}_{-0.02} $ &1209/1042\\
B3& $0.06^{+0.02}_{-0.03}$&$0.24^{+0.08}_{-0.06} $&$3.94^{+0.78}_{-1.07}$&$1.67^{+0.01}_{-0.01} $ &$3^{\rm f}$ &$2.59^{+0.17}_{-0.26} $ &$<0.05 $ &$<1.65$ &1131/986\\
B4 & $0.07^{+0.02}_{-0.04}$&$0.26^{+0.04}_{-0.08} $&$8.87^{+2.52}_{-1.13}$&$1.71^{+0.01}_{-0.02} $ &$3^{\rm f}$ &$3.19^{+0.20}_{-0.24} $ &$<0.05$ &$<0.56$ &1121/1004\\
B5&$0.07^{+0.02}_{-0.03}$&$0.23^{+0.07}_{-0.05} $&$5.99^{+0.76}_{-0.61}$&$1.68^{+0.02}_{-0.01} $ &$3^{\rm f}$ &$3.01^{+0.26}_{-0.33} $ &$<0.04 $ &$<0.08$ &1079/867\\
\hline
\end{tabular}
\leftline{The errors are quoted for 90\% confidence level (1.6 $\sigma$). $^{\rm f}$ indicate that the parameter is fixed during the fitting.}
\leftline{\nh is the hydrogen column density in $10^{22}$ \pcm. $T_{\rm in}$ and $N_{\rm DBB}$ are the inner disc temperature (in keV) and normalization of {\tt DISKBB} model, respectively.}
\leftline{$\Gamma$, $N_{\rm nthcomp}$ are the photon index and normalization of {\tt NTHCOMP} model, respectively. $E_{\rm edge}$ is the absorption edge (in keV) of the {\tt EDGE} model.}
\leftline{$L_{\rm X}$ is the broadband X-ray luminosity (in $10^{38}$ \eps) in the energy range of $0.1-100$~keV. $f_{\rm disc}$ is the thermal fraction and defined as $f_{\rm disc} = F_{\rm disc}/F_{\rm tot}$,}
\leftline{where $F_{\rm disc}$ and $F_{\rm tot}$ are disc flux and total flux, respectively.}
\leftline{$a^*$, $M_{\rm BH}$, $\dot{\rm M}$, $\kappa$ are the dimensionless spin parameter, mass of the black hole (in $M_{\odot}$), mass accretion rate (in $10^{18}$ \gps), and spectral}
\leftline{hardening factor of {\tt KERRBB} model respectively.}
\leftline{$\Gamma$ and $f_{\rm Scat}$ are the photon index and scattering fraction of {\tt SIMPL} model, respectively.}
\leftline{${\rm \xi}$, $R_{\rm refl}$ and $N_{\rm rel}$ are the ionization parameter (in \ecps), reflection fraction and normalization (in $10^{-3}$ \phc) of {\tt RELXILL} model,}
\leftline{respectively.}
\end{table*}

\section{Discussion}
\label{sec:discussion}
We carried out timing and spectral analysis of LMC~X--1 and LMC~X--3 using data from multi-epoch simultaneous observations with \nustar and \swift observatories. As both the sources are weak in hard X-rays, data from the \swift/XRT and \nustar were used in $0.5-30$~keV range in our analysis. Simultaneous observations at three epochs between 2014 and 2016 for LMC~X--1 and five epochs between 2015 and 2019 for LMC~X--3 are analyzed in the present work.

LMC~X--1 has always been found to be in the thermal dominated HSS \citep{Nowak2001}, whereas LMC~X--3 shows state transitions \citep{Torpin2017}. However, LMC~X--1 and LMC~X--3 were caught in the HSS during all the epochs of observations used in the present work. Using background subtracted light curves, we estimated the fractional rms variabilities of both the sources during the observations and found to be $<2$\%. The HID of individual observations did not show any systematic variation. The power in the PDS got diminished above $\sim 10$~Hz. We did not detect any signature of QPOs in the PDS of both sources during all the observations. The observed timing properties of both the sources are found to be similar to that during typical HSS of BHs \citep{RM06}.

Spectral analysis of data from all three observations of LMC~X--1 revealed that the emission was dominated by thermal disc flux ($f_{\rm disc} > 0.85.$) with low scattering fraction ($f_{\rm Scat} \sim 0.1$). This indicates insubstantial emission from Compton Corona. The reflection  was found to be low with $R_{\rm refl}<0.1$ during all three epochs of observations. This could be associated with the fact that the Comptonized flux was very low. LMC~X--3 also showed similar behaviour during the observations at all five epochs. We did not find any evidence of the reflection component in the $0.5-30$~keV spectrum from any of the five observations with $R_{\rm refl}<0.06$. This confirms that LMC~X--3 was in the HSS during the observations used in the present work, though state transition has been observed in the past.

Spectral fitting of data from \swift/XRT and \nustar simultaneous observations allowed us to constrain the mass and spin of LMC~X--1 and LMC~X--3. The estimated mass of both the BHs agrees well with the earlier reported values. LMC~X--1 is reported to be a high spinning BH. \citet{Gou2009} estimated the spin of LMC~X--1 as $a^* = 0.92^{+0.05}_{-0.07}$ from {\tt RXTE} observations. \citet{Steiner2012} estimated the spin as $a^*=0.97^{+0.02}_{-0.25}$ from the reflection spectroscopy. \citet{Mudambi2020} and \citet{Bhuvana2021} estimated the spin of LMC~X--1 as $a^*=0.93^{+0.02}_{-0.02}$ and $a^* \sim 0.82-0.92$, respectively, from the {\it AstroSat} observations. In this work, we obtained the spin of LMC~X--1 in the range of $a^*\sim 0.92-0.95$ from \nustar observations. Our estimated values of the spin matches well with the previous observations. On the other hand, LMC~X--3 is reported to host a low spinning BH with spin parameters, $a^{*}=0.25^{+0.20}_{-0.29}$ \citep{Steiner2014}. Recently, {\it AstroSat} observations revealed the spin of LMC~X--3 in the range of $a^* \sim 0.22-0.41$ \citep{Bhuvana2021}. Our estimated value of the spin of the BH is in the range of $a^{*} \sim 0.19-0.29$.

From the {\tt diskbb} normalization ($N_{\rm DBB}$), we can calculate the inner edge of the accretion disc. The disc normalization is given by, $N_{\rm DBB} = (r_{\rm in}^2/D_{\rm 10}^2) \cos i$, where $D_{\rm 10}$ is the distance of the source in the unit of 10~kpc, $r_{\rm in}$ is apparent inner accretion disc radius in km, and $i$ is the inclination angle \citep{Mitsuda1984,Makishima1986}. The inner edge of the disc is given by, $R_{\rm in}=x \kappa^2 r_{\rm in}^2$, where $x=0.41$ is the correction factor \citep{Kubota1998} and $\kappa$ is the spectral hardening factor \citep{Shimura-Takahara1995}. From the spectral fittings, we observed that the inner edge of the disc for LMC~X--1 varied in the range of $R_{\rm in} \sim 40-43$~km $\sim 2.5$~$R_{\rm g}$, considering $\kappa=1.55$ (see Sec.~\ref{sec:lmc-x1}).
For LMC~X--3, we obtained the inner edge of the disc in the range of, $R_{\rm in} \sim 47-53$~km $\sim 4.5-5$~$R_{\rm g}$ for different epochs, considering  $\kappa=1.7$ (see Sec.~\ref{sec:lmc-x3}). Considering the spin, $a* \sim 0.9$, the ISCO is located at $R_{\rm ISCO} \sim 2.3$~$R_{\rm g}$ for LMC~X--1. For LMC~X--3, the ISCO is located at $R_{\rm ISCO} \sim 5$~$R_{\rm g}$, considering spin $a^* \sim 0.25$. Hence, for both LMC~X--1, and LMC~X--3, the inner edge of the disc ($R_{\rm in}$) is obtained to be consistent with the $R_{\rm ISCO}$.

We obtained the broad-band X-ray luminosity ($L_{\rm X}$) in the energy range of $0.1-100$~keV for LMC~X--1 and LMC~X--3. The luminosity is obtained to be in the range of $L_{\rm X} \sim 1.38-1.52 \times 10^{38}$ \eps, which corresponds to the Eddington ratio, $L_{\rm X}/L_{\rm Edd} \sim 0.1-0.11$ for LMC~X--1. Here, $L_{\rm Edd}$ is the Eddington luminosity and given by, $L_{\rm Edd} = 1.26 ~\times~10^{38} ~ (M_{\rm BH}/M_{\sun})$ \eps. During our observations, the luminosity of LMC~X--3 was estimated to be $L_{\rm X} \sim 1.46-4.06 ~\times~ 10^{38}$ \eps. The Eddington ratio for LMC~X--3 was, $L/L_{\rm Edd} \sim 0.16-0.46$. From this, we conclude that during our observation period, LMC~X--1 was stable while LMC~X--3 showed variation in the mass accretion rate.

The accretion efficiency is given by, $\eta = L_{\rm X}/\dot{\rm M}c^2$, where $c$ is the speed of light. The mass accretion rate for LMC~X--1 and LMC~X--3 are derived to be $\dot{\rm M} \sim 1.06-1.31 \times 10^{18}$ \gps, and $\dot{\rm M} \sim 4.05-9.24 \times 10^{18}$ \gps, respectively. The accretion efficiency are obtained to be $\eta \sim 0.13$ and $\eta \sim 0.04$ for LMC~X--1 and LMC~X--3, respectively. The accretion efficiency is higher for LMC~X--1. This is expected as LMC~X--1 is a high spinning BH, while LMC~X--3 is a low spinning BH.

\section{Summary}
We studied two extra-galactic black hole X-ray binaries LMC~X--1 and LMC~X--3 using data from simultaneous observations with the \swift and \nustar observatories. Following are the key findings from our analysis.

\begin{enumerate}
\item Spectral and timing analysis of \swift/XRT and \nustar data revealed that both LMC~X--1 and LMC~X--3 were in the HSS during all the observations.
\item We did not find any signature of QPO in the PDS of both the sources. Very weak variability was also observed in the PDS of each source with fractional rms amplitude $<2$\%.
\item The spectra of both the sources were associated with a disc of temperature $T_{\rm in} \sim 1$~keV and photon index, $\Gamma >2.4$. The thermal disc fraction was observed as $f_{\rm disc} \geq 80$\% with a small scattering fraction, $f_{\rm Scat} \sim 0.1$.
\item The Fe K$\alpha$ emission line was detected in the spectra of LMC~X--1, whereas no such feature was present in the spectra of LMC~X--3.
\item The spin of LMC~X--1 and LMC~X--3 were estimated to be $0.92-0.95$ and $0.19-0.29$, respectively.
\item The broadband X-ray luminosities were observed to be in the range of $L_{\rm X} \sim 1.38-1.52 \times 10^{38}$ \eps, and $L_{\rm X} \sim 1.46-4.06 ~\times~ 10^{38}$ \eps, for LMC~X--1 and LMC~X--3, respectively. This corresponds to the Eddington ratio of $\sim 0.1-0.11$ and $\sim 0.16-0.46$ for LMC~X--1 and LMC~X--3, respectively.
\item The values of accretion efficiency was obtained to be $\eta \sim 0.13$ and $\eta \sim 0.04$ for LMC~X--1 and LMC~X--3, respectively.
\end{enumerate}

\section*{Acknowledgements}
We acknowledge the anonymous reviewer for the helpful comments and suggestions which improved the paper. Work at Physical Research Laboratory, Ahmedabad, is funded by the Department of Space, Government of India. This research has made use of the {\it NuSTAR} Data Analysis Software ({\tt NuSTARDAS}) jointly developed by the ASI Space Science Data Center (SSDC, Italy) and the California Institute of Technology (Caltech, USA). This work was made use of XRT data supplied by the UK Swift Science Data Centre at the University of Leicester, UK.

\section*{DATA AVAILABILITY}
We used archival data of {\it Swift}, and {\it NuSTAR} observatories for this work.

\bibliographystyle{mnras}
\bibliography{ms_lmc}



\appendix



\bsp	
\label{lastpage}
\end{document}